## Title page

Type of manuscript: Original Research

Manuscript title: Convergence fragility in probit Bayesian kernel machine regression implemented in the *bkmr* R package for binary-outcome environmental mixture analyses: a simulation study

Authors and affiliations:

Akifumi Eguchi[1*], Takayuki Kawashima[2], Tomotaka Momozaki[3], Tomoyuki Nakagawa[4]

[1] Center for Preventive Medical Sciences, Chiba University, Chiba, Japan

[2] School of Computing, Institute of Science Tokyo, Tokyo, Japan

[3] Department of Information Science and Technology, Tokyo University of Science

[4] School of Data Science, Meisei University, Tokyo, Japan

*Corresponding author:

Akifumi Eguchi

Center for Preventive Medical Sciences, Chiba University

Inage-ku Yayoi-cho 1-33, Chiba, Japan

Telephone: +81-43-290-3896

Email: a_eguchi@chiba-u.jp

ORCID: 0000-0001-5858-8132

Running head: Probit BKMR convergence diagnostics

Conflicts of Interest: The authors declare no conflicts of interest.

Source of Funding: This work was supported by the Environment Research and Technology Development Fund (JPMEERF20245M01) of the Environmental Restoration and Conservation Agency provided by the Ministry of the Environment of Japan. The authors did not receive funding specifically to support open access publication of this article.

Data and code availability: Supplemental Digital Content documents the simulation task table, numerical completion status, diagnostic settings, task-level summaries, and parameter-level diagnostics. The analysis code and reproduction materials will be made publicly available in a GitHub repository after the preprint is posted.

Until then, the scripts, task tables, summary outputs, parameter-level diagnostic CSVs, and environment information are available from the corresponding author on request.

Acknowledgements: We thank the developers of the *bkmr* and posterior R packages for making their software publicly available.

Declaration on AI-assisted editing: ChatGPT 5.5 Pro and Claude Opus 4.8 were used only to assist with language editing and proofreading. The authors reviewed and verified all AI-assisted edits and take full responsibility for the scientific content, analyses, interpretations, and conclusions of the manuscript.

Author contributions (CRediT): A.E.: Conceptualization, Formal analysis, Funding acquisition, Methodology, Software, Writing - original draft, Writing - review & editing. T.K.: Formal analysis, Methodology, Writing - review & editing. T.M.: Methodology, Writing - review & editing. T.N.: Writing - review & editing.

## Abstract

Background. Bayesian kernel machine regression (BKMR) is widely used for exposure-mixture analyses with binary outcomes through a probit extension. Because a *bkmr* fit can complete without providing adequate effective posterior information, simulation studies should separate execution success from MCMC convergence diagnostics.

Methods. We evaluated the public *bkmr* probit workflow using *bkmr*::SimData() for data generation, *bkmr*::kmbayes() for model fitting, and posterior for convergence diagnostics. The balanced generator used family = "binomial", hfun = 2, beta.true = 0.5, ind = 1:2, and M = 4. SimData() generated the covariate as X = 3*cos(z1) + 2*rnorm(n). Four chains were initialized with chain-specific randomized starting values generated reproducibly from the fixed initial-value base seed 20260621. These values affected only the initial state of the sampler and did not alter the BKMR model, default priors, or Metropolis-Hastings proposals.

Results. Of 431 prespecified tasks, 430 returned fitted objects and one task had a numerical non-completion. Diagnostic adequacy was limited: rank-normalized R-hat <= 1.01 threshold was achieved in 55/431 tasks, bulk-ESS >= 400 in 85/431, tail-ESS >= 400 in 44/431, and both ESS criteria in 44/431. The primary diagnostic criterion, R-hat at or below the 1.01 threshold with both bulk-ESS and tail-ESS >= 400, was met in 30/431 prespecified tasks, corresponding to 30/430 completed fits.

Conclusions. Completion of probit BKMR fits in *bkmr* should not be equated with convergence of the retained MCMC draws. Applied analyses should report the number of chains, warmup and retained iterations, rank-normalized R-hat, bulk-ESS, and tail-ESS rather than rely on a fixed iteration count or on fit completion alone.

## What this study adds

This simulation study evaluates the public bkmr probit workflow using posterior-based MCMC diagnostics. Of 431 prespecified tasks, 430 returned fitted objects and one had a numerical non-completion. Only a small minority of tasks met the prespecified criteria based on rank-normalized R-hat, bulk-ESS, and tail-ESS, following the recommendations of Vehtari et al. The study therefore supports routine reporting of both model-fitting completion and posterior diagnostic adequacy in binary-outcome BKMR mixture analyses.

## 1. Introduction

Estimating the joint health effects of environmental exposure mixtures is a central problem in environmental epidemiology. Bayesian kernel machine regression (BKMR) has been widely adopted because it accommodates nonlinear and non-additive exposure-response surfaces, variable selection, and correlated exposures.[1–3] The R package *bkmr* extends BKMR to binary outcomes through a probit formulation with latent-normal data augmentation, following the Albert-Chib scheme.[2,4]

Standard MCMC practice requires multiple chains together with diagnostics that assess both between-chain agreement and the amount of independent posterior information.[5,6] Vehtari et al. recommend improved R-hat diagnostics based on rank normalization and folding, together with effective sample-size diagnostics for the bulk and tails of the posterior distribution.[6] We used these diagnostics because traditional R-hat can fail to flag some forms of nonconvergence, whereas rank-normalized and folded R-hat together with bulk-ESS and tail-ESS provide a more robust assessment of the retained draws; these diagnostics are also implemented in widely used Bayesian software such as Stan.[6,7] These diagnostics are particularly relevant for BKMR because posterior inclusion probabilities, posterior mean effect estimates, and exposure-response summaries are meaningful only when the monitored chains provide adequate posterior information.

The probit BKMR vignette for *bkmr* also notes that visual inspection of trace plots may indicate the need for more than 10,000 iterations, and recommends longer runs, for example 50,000 iterations, with convergence assessment before presenting final results.[8] Concerns about BKMR convergence and diagnostic reporting have been raised previously. Duttweiler et al., in a paper developing general-purpose diagnostics for difficult MCMC settings, used probit BKMR as a working example after multiple investigators reported convergence problems with the *bkmr* probit algorithm, and described a lambda-proposal modification that was subsequently incorporated into the development version of *bkmr*.[9,10] Separately, Hasan et al. reported reliability concerns for Gaussian BKMR under skewed and correlated exposure data, including type-I error inflation as the coefficient of variation increased and limitations of the conventional posterior inclusion probability threshold. Their study did not examine the probit implementation.[11]

When applied BKMR reports provide iteration counts without accompanying rank-normalized R-hat, bulk-ESS, tail-ESS, or related diagnostics, readers cannot assess whether the reported posterior summaries are supported by adequately mixed chains. This is consequential because Monte Carlo estimates of posterior expectations are justified by Markov chain central limit theorem conditions, for which geometric ergodicity is a common sufficient condition. Without empirical evidence from multi-chain diagnostics and adequate effective sample sizes, Monte Carlo error in reported posterior means, intervals, and posterior inclusion probabilities is difficult to evaluate; a small R-hat alone does not establish adequate posterior computation if bulk-ESS or tail-ESS remains insufficient.[6,12]

The present study addresses a focused implementation question relevant to applied mixture analyses: when the public *bkmr* probit workflow is run in a controlled simulation setting, how often do completed fits satisfy contemporary MCMC diagnostic criteria?[6] We used the public *bkmr* workflow available to applied analysts and evaluated fit completion separately from diagnostic adequacy of the retained MCMC draws. Specifically, we examined rank-normalized R-hat, bulk effective sample size, and tail effective sample size across prespecified tasks, including sample-size sweeps, variable-selection settings, longer-chain checks, and default-generator support checks.

## 2. Methods

### 2.1 Simulation setting and task structure

The main balanced simulation used *bkmr*::SimData(n = n, M = 4, family = "binomial", hfun = 2, beta.true = 0.5, ind = 1:2). The hfun = 2 surface makes two exposure components active. In SimData(), the exposure matrix Z is generated first and the single covariate is then generated as X = 3*cos(z1) + 2*rnorm(n); therefore X is not an independently generated standard-normal covariate.[10] We use the term balanced generator to denote this outcome-distribution setting: it was chosen to produce less imbalanced binary outcomes than the default-generator support setting with hfun = 3 and beta.true = 2, while retaining two active exposure components.

The task table contained 431 model-fitting tasks; the exact task-table strata are shown in Supplemental Digital Content eTable 2.1. The balanced main sweep used K = 50 independent dataset realizations at n = 40, 80, 160, and 320. In this task table, the n = 40 balanced main scenario used varsel = TRUE, whereas n = 80, 160, and 320 used varsel = FALSE. A balanced n = 40 ablation with varsel = FALSE was run using the same dataset-seed structure. Additional tasks included balanced long-run checks at 50,000 and 100,000 iterations, a balanced single-seed long-run check at n = 40, default-generator support checks with hfun = 3 and beta.true = 2, and public-verification tasks. Unless otherwise noted, four parallel chains were used, with warmup equal to half of the total iteration count.

### 2.2 Model fitting and starting values

All models were fitted with *bkmr*::kmbayes(y = y, Z = Z, X = X, family = "binomial", est.h = FALSE) and the scenario-specific varsel setting. Four chains were initialized with chain-specific randomized starting values generated reproducibly from the fixed initial-value base seed 20260621. The four chain seeds (142709412, 1107285817, 1907135512, 672667230) seeded the sampler for each chain, while the chain-specific dispersed starting values were drawn from a task-specific seed derived from the fixed initial-value base seed 20260621. The initial-value distributions were selected to provide dispersed starting points while respecting the parameter constraints of the probit BKMR implementation. These values affected only the initial state of the sampler.

We retained the default *bkmr*::kmbayes() control parameters. In the binomial/probit model, sigsq.eps is fixed to 1. The default prior for lambda is parameterized by mu.lambda = 10 and sigma.lambda = 10, with a gamma Metropolis-Hastings proposal using lambda.jump = sqrt(10) for binomial fits. The default r prior is inverse-uniform with r.a = 0 and r.b = 100, and r is updated by Metropolis-Hastings using the package default proposal scales. When variable selection was enabled, the default beta-binomial prior on the inclusion indicators used a.p0 = b.p0 = 1.9 We did not modify the BKMR likelihood, kernel, prior distributions, or Metropolis-Hastings proposal mechanisms. The starting-value distributions and full prior and proposal details are given in eAppendix 1 and eTables 1.1 and 1.2.

### 2.3 Convergence diagnostics and thresholds

For each task, we retained post-warmup draws from four chains and computed diagnostics with the posterior package. posterior::rhat() returns the maximum of rank-normalized split R-hat and rank-normalized folded-split R-hat.[6,13] We also computed bulk-ESS and tail-ESS. Bulk-ESS and tail-ESS were computed from rank-normalized split chains; because the four original chains were split into first and second halves, the ESS threshold of 400 corresponds to an average of 50 effective draws per split chain. The monitored stochastic quantities were lambda, beta[1], and r[z1] through r[z4]. In the probit model, sigsq.eps is fixed at 1 as the usual latent-scale identification constraint and is therefore not a monitored stochastic parameter; its R-hat and ESS are undefined and were not interpreted as evidence of nonconvergence.[4]

Following Vehtari et al., the primary diagnostic criterion was no monitored stochastic parameter above the 1.01 R-hat threshold and no monitored stochastic parameter with either bulk-ESS or tail-ESS below 400.6 In the CSV implementation, a posterior-parameter row was flagged when R-hat > 1.01 or when either ESS measure was < 400.

## 3. Results

### 3.1 Fit completion and overall diagnostic adequacy

Of 431 prespecified tasks, 430 returned fitted objects and one task had a numerical non-completion. The non-completion occurred in the balanced n = 320, varsel = FALSE scenario and returned the error message "H must be positive definite". This model-fitting non-completion should be interpreted separately from MCMC diagnostic adequacy among completed fits. Under the primary diagnostic criterion, 30/431 prespecified tasks, corresponding to 30/430 completed fits, passed R-hat and both ESS thresholds.

Table 1 summarizes fit completion, numerical non-completion, and diagnostic pass counts across all 431 prespecified tasks.

**Table 1. Overall probit BKMR completion and posterior diagnostic summary.**

| Diagnostic summary | Count among 431 tasks | Interpretation |
| --- | --- | --- |

| Fit completed | 430/431 | Execution status only |
|---|---|---|
| Numerical non-completion | 1/431 | Model-fitting non-completion; not a diagnostic failure |
| R-hat <= 1.01, considered alone | 55/431 | Vehtari-style R-hat threshold |
| bulk-ESS >= 400, considered alone | 85/431 | Bulk posterior information |
| tail-ESS >= 400, considered alone | 44/431 | Tail posterior information |
| bulk-ESS and tail-ESS >= 400 | 44/431 | Both ESS criteria |
| R-hat <= 1.01 plus both ESS criteria | 30/431 | Primary diagnostic pass |
| Primary diagnostic pass among completed fits | 30/430 | Primary pass denominator limited to completed fits |

### 3.2 Balanced generator and n = 40 ablation

Under the balanced generator, all K = 50 runs completed for each scenario shown in Table 2 except for one numerical non-completion at n = 320 with varsel = FALSE. The number of runs meeting the primary diagnostic criterion increased with sample size in the varsel = FALSE sequence from n = 80 to n = 320, but many completed fits remained flagged, especially by ESS. At n = 40, the varsel = TRUE main scenario had 1/50 primary diagnostic pass and the varsel = FALSE ablation had 0/50 primary diagnostic passes, showing that toggling variable selection did not by itself resolve the diagnostic limitations at the smallest sample size.

**Table 2. Balanced generator scenario diagnostics. The primary pass requires R-hat at the 1.01 threshold and both bulk-ESS and tail-ESS >= 400.**

| Scenario | Completed | Median (worst) R-hat | Median (worst) bulk-ESS | Median (worst) tail-ESS | Primary pass |
|---|---|---|---|---|---|
| n=40; varsel=TRUE | 50/50 | 1.043 (1.859) | 116.4 (5.7) | 48.0 (11.4) | 1/50 |
| n=80; varsel=FALSE | 50/50 | 1.035 (2.043) | 126.7 (5.3) | 53.7 (11.1) | 1/50 |
| n=160; varsel=FALSE | 50/50 | 1.013 (1.593) | 417.9 (6.8) | 243.2 (13.4) | 10/50 |
| n=320; varsel=FALSE | 49/50 | 1.012 (1.578) | 461.2 (6.9) | 322.5 (11.5) | 11/50 |

| n=40; varsel=FALSE ablation | 50/50 | 1.051 (1.586) | 80.9 (6.9) | 44.9 (10.9) | 0/50 |
|---|---|---|---|---|---|

### 3.3 Longer chains and default-generator support checks

Table 3 summarizes the longer-chain balanced checks and the default-generator support checks.

**Table 3. Longer-chain balanced checks and default-generator support checks.**

| Check | Completed | Primary pass | Median (worst) R-hat | Median (worst) bulk-ESS | Median (worst) tail-ESS |
|---|---|---|---|---|---|
| Balanced n=40, iter=50000 | 3/3 | 2/3 | 1.009 (1.011) | 748.6 (365.1) | 797.8 (345.5) |
| Balanced n=40, iter=100000 | 3/3 | 2/3 | 1.006 (1.313) | 785.9 (9.9) | 781.1 (15.3) |
| Balanced n=80, iter=50000 | 3/3 | 1/3 | 1.004 (1.293) | 1002.1 (10.6) | 394.6 (11.5) |
| Balanced n=80, iter=100000 | 3/3 | 2/3 | 1.002 (1.575) | 2552.2 (6.8) | 1010.8 (11.3) |
| Default n40_varsel_true | 20/20 | 0/20 | 1.670 (2.917) | 6.5 (4.6) | 14.3 (11.4) |
| Default n80_varsel_false | 20/20 | 0/20 | 1.361 (2.124) | 9.7 (5.3) | 22.6 (11.5) |

Longer chains improved some diagnostics but did not uniformly resolve them. In the balanced three-seed long-run, n = 40 had 2/3 primary passes at both 50,000 and 100,000 iterations, whereas n = 80 had 1/3 primary pass at 50,000 iterations and 2/3 primary passes at 100,000 iterations. The default-generator support checks, based on hfun = 3 and beta.true = 2 at n = 40 with varsel = TRUE and n = 80 with varsel = FALSE, were more difficult: all 40 tasks completed, but none met the primary diagnostic criterion. Consistent with Vehtari et al., a failure of ESS to grow sufficiently with more draws indicates that simply running longer chains need not resolve convergence issues; in the default-generator long-run checks in Supplemental Digital Content eTable 4.3, no three-seed long-run task met the primary criterion even at 100,000 iterations, and ESS remained below the prespecified threshold.6

### 3.4 Parameter-level patterns

The reproduction outputs included one parameter-level diagnostic CSV for each completed fit. These 430 outputs were summarized by parameter group in Supplemental Digital Content eTable 5.1. Diagnostic flags were not confined to a single parameter group. beta and lambda frequently showed elevated rank-normalized R-hat or low bulk-ESS or tail-ESS. The r parameters had acceptable median diagnostics, but many tasks had at least one flagged r parameter because four r parameters were monitored per task.

## 4. Discussion

In this simulation study, 430 of 431 prespecified probit BKMR fitting tasks returned fitted objects, and one task had a numerical non-completion. Successful completion did not imply adequate posterior diagnostics. Using the primary diagnostic criterion based on the recommendations of Vehtari et al.—rank-normalized R-hat <= 1.01, bulk-ESS >= 400, and tail-ESS >= 400—only 30 of 431 prespecified tasks, or 30 of 430 completed fits, satisfied all diagnostic thresholds. This result should therefore be read as a diagnostic adequacy finding among completed public *bkmr* fits, not as a numerical failure rate for kmbayes().[6,12]

This distinction is relevant for applied mixture analyses because posterior inclusion probabilities, exposure-response summaries, and mixture-effect estimates are interpretable only to the extent that the retained draws provide adequate posterior information. In that sense, a returned model object is not sufficient evidence that posterior summaries are reliable.[1,2,6]

These findings extend prior concerns about probit BKMR computation. Duttweiler et al. used probit BKMR as a working example in methodological work on general MCMC diagnostics and described lambda-proposal changes; the corresponding modification is recorded in the development version of *bkmr*.[8–10] Our results do not contradict that implementation improvement; rather, they show that, even when numerical fitting completes, posterior diagnostic adequacy still needs to be evaluated directly. The chain-specific randomized starting values used in the present simulations were introduced to make the probit initialization reproducible and independent of the internal probit GLM starting step while providing dispersed initial states. They did not modify the BKMR likelihood, kernel, prior distributions, or Metropolis-Hastings proposal mechanisms. The results therefore describe the public *bkmr* sampler under its default inferential settings, not a new BKMR algorithm or a modified posterior target.

A closer look at the individual diagnostic components showed that R-hat alone did not provide sufficient evidence of diagnostic adequacy. Although 55 tasks met the primary 1.01 R-hat threshold when considered alone, only 30 tasks also satisfied both bulk-ESS and tail-ESS criteria. Thus, reporting that R-hat was below a permissive threshold would not be sufficient to support posterior inference if bulk-ESS or tail-ESS remained too small. Conversely, a fit should not be described as converged merely because kmbayes() returned an object without error.[5,6,12]

The longer-chain checks also support this point. Increasing the number of iterations improved some diagnostics, but the improvement was not uniform across scenarios or diagnostic components. In particular, some runs showed acceptable R-hat or bulk-ESS while tail-ESS remained below the prespecified threshold. This pattern supports reporting both bulk-ESS and tail-ESS rather than relying on R-hat alone or on a single effective-sample-size summary.[6,13]

Parameter-level summaries in Supplemental Digital Content eTable 5.1 showed that diagnostic flags were not confined to a single model component. beta and lambda frequently showed elevated rank-normalized R-hat or low bulk-ESS or tail-ESS. As shown in Supplemental Digital Content eTable 5.1, the r parameters showed acceptable median diagnostics when summarized across parameter rows, but task-level flags were frequent because four r parameters, r[z1] through r[z4], were monitored for each completed fit and a task was flagged if any monitored parameter failed the diagnostic criteria. This parameter-level summary therefore helps explain why task-level diagnostic failures were common even when some parameter-group medians appeared acceptable.

These findings support a practical reporting recommendation. Probit BKMR analyses should report the number of chains, total and warmup iterations, retained draws, rank-normalized R-hat, bulk-ESS, and tail-ESS for the monitored posterior parameters. Reports should state whether diagnostic criteria were prespecified and should clearly separate model execution from convergence evidence. Fit completion, iteration count, and a returned model object are not sufficient evidence that posterior summaries are reliable.[6,12]

Several limitations should be noted. The simulations cover a finite set of controlled balanced and default-generator scenarios rather than all possible exposure distributions, signal strengths, outcome prevalences, or variable-selection configurations. The aggregate diagnostic-pass proportion across 431 tasks is determined by the composition of the prespecified task table and should not be interpreted as the prevalence of diagnostic concerns in all applied BKMR analyses. The analysis evaluates the public *bkmr* workflow and does not propose or validate a replacement sampler. The chain-specific randomized starting values improved reproducibility and provided dispersed initial states, but they do not rule out numerical or diagnostic problems under other initial-value distributions, seeds, or data configurations. One numerical non-completion was observed, but the present analysis did not attempt to localize the internal source of that error.

Finally, the diagnostic thresholds used here identify insufficient evidence for reliable posterior computation under the specified criteria; they do not prove convergence for tasks that pass or rule out other Monte Carlo error concerns for every posterior estimand.[6,12]

In summary, this simulation study of 431 prespecified probit BKMR tasks found 430 completed fits, one numerical non-completion, and frequent posterior diagnostic flags among completed fits. Applied binary-outcome BKMR mixture analyses should not rely on fit completion or iteration count alone; they should report rank-normalized R-hat, bulk-ESS, and tail-ESS, preferably with parameter-level summaries that identify which model components contribute to diagnostic concerns.

## Supplemental Digital Content (eAppendix)

Convergence fragility in probit Bayesian kernel machine regression for binary-outcome environmental mixture analyses: a simulation study

This Supplemental Digital Content documents the *bkmr* simulation analysis reported in the main manuscript. Model fitting used *bkmr*::kmbayes(), and MCMC diagnostics were computed with posterior. Specifically, posterior::rhat() was used for the maximum of rank-normalized split R-hat and rank-normalized folded-split R-hat, and posterior::ess_bulk() and posterior::ess_tail() were used for bulk-ESS and tail-ESS. The supplement provides task-level and parameter-level diagnostics from the reproduction outputs. These ESS diagnostics were computed from rank-normalized split chains; with four original chains split into first and second halves, ESS >= 400 corresponds to an average of at least 50 effective draws per split chain.

# eAppendix 1. Data generation, fitting, and default sampler settings

## 1.1 Data generation

Balanced datasets were generated by *bkmr*::SimData(n = n, M = 4, family = "binomial", hfun = 2, beta.true = 0.5, ind = 1:2). The default-generator support checks used hfun = 3 and beta.true = 2, also with M = 4. In SimData(), the generated covariate is X = 3*cos(z1) + 2*rnorm(n). Thus, the covariate adjusted for in the fitted model should not be described as an independently generated standard-normal covariate.

## 1.2 Model fitting and initial values

Each task fitted four *bkmr*::kmbayes() chains with family = "binomial" and est.h = FALSE. Initial values were generated reproducibly from the fixed initial-value base seed 20260621. The four chain seeds were 142709412, 1107285817, 1907135512, and 672667230, and an initial_value_task_seed was deterministically derived for each task. The starting-value distributions were h.hat ~ Uniform(-2, 2), beta ~ Uniform(-1, 1), positive r and |ystar| generated as exp(Uniform(-1, 1)), and lambda ~ log-Uniform(1, 20). The sign of ystar was constrained by the observed binary outcome y, sigsq.eps was fixed at 1, and, for varsel = TRUE, delta ~ Bernoulli(0.5) with r = 0 where delta = 0. These values were initialization choices only. No alternative BKMR model, prior distribution, or MH proposal was introduced.

**eTable 1.1. Explicit starting.values supplied to kmbayes().**

| Quantity | Value used in the present analysis | Role |
| --- | --- | --- |
| initial_value_seed | 20260621 | Fixed base seed for reproducible initial-value generation |
| chain seeds | 142709412; 1107285817; 1907135512; 672667230 | Per-chain seeds for the MCMC sampler (set immediately before each kmbayes() chain); not used to generate the starting values |
| initial_value_task_seed | Deterministically derived for each task from the base seed | Task-specific reproducible seed from which the four chains' dispersed starting values are drawn |
| h.hat | Uniform(-2, 2) for each observation | Initial h state |
| beta | Uniform(-1, 1) for the single covariate coefficient | Initial fixed-effect coefficient |
| sigsq.eps | 1 | Fixed residual variance in probit family |

| r | exp(Uniform(-1, 1)) for positive r; r = 0 when varsel = TRUE and delta = 0 | Initial kernel-bandwidth vector respecting positivity and variable-selection constraints |
|---|---|---|
| lambda | log-Uniform(1, 20) | Initial variance-ratio parameter |
| delta | Bernoulli(0.5) when varsel = TRUE | Initial variable-selection indicators |
| ystar | sign constrained by y, with \|ystar\| = exp(Uniform(-1, 1)) | Initial latent probit outcome |

### 1.3 Default *bkmr* prior distributions and MCMC update/proposal settings

All analyses used *bkmr*::kmbayes() with family = "binomial" and retained the default control.params unless explicitly stated otherwise. The explicit starting.values described in the main text were used only to set the initial state of the Markov chain and did not modify the prior distributions, proposal mechanisms, or conditional updates summarized below.

eTable 1.2. Default prior distributions and MCMC update/proposal settings retained from *bkmr*::kmbayes().

| **Item** | **Default prior / update** | **Default update/proposal setting** | **Use in the present analysis** |
|---|---|---|---|
| lambda | Gamma prior, parameterized in *bkmr* by mu.lambda = 10 and sigma.lambda = 10; shape = 1 and rate = 0.1. | Gamma Metropolis-Hastings proposal using the package lambda update; for binomial/probit fits, lambda.jump = sqrt(10). | Retained without override. Monitored as lambda; MH acceptance summarized as acc.lambda. |
| r | r.prior = "invunif", r.a = 0, and r.b = 100; equivalently, 1/r ~ Uniform(0, 100), so r >= 0.01. | For non-variable-selection r updates, truncated-normal MH proposal on r over [1/r.b, 1/r.a], with r.jump = 0.1. | Retained without override. Monitored as r[z1] through r[z4]; MH acceptance summarized as acc.r. |
| delta | When varsel = TRUE, beta-binomial model-size prior with a.p0 = 1 and b.p0 = 1. | Component-wise add/drop moves toggle one exposure component. Within-model r updates use the r proposal with r.jump2 = 0.1. | Retained without override for variable-selection scenarios. delta itself was not summarized by R-hat/ESS in the main task-level table; rdelta_move acceptance was recorded descriptively. |
| beta | Fixed-effect coefficients updated from the full conditional normal distribution; no user-specified beta prior or prior precision override was supplied. | Conditional normal/Gibbs update rather than a Metropolis-Hastings proposal. | Retained without override. Monitored as beta[1]. |
| sigsq.eps | For family = "binomial", *bkmr* fixes sigsq.eps = 1. | Not updated in binomial/probit models. | Fixed at 1 throughout. R-hat and ESS are undefined for this constant quantity and were not interpreted as convergence failures. |
| ystar | Latent probit outcome used for the Albert-Chib data-augmentation representation. | With est.h = FALSE, updated from a truncated multivariate normal distribution using tmvtnorm::rtmvnorm() conditional on | Updated internally by *bkmr*. Not treated as a target posterior |

| Item | Default prior / update | Default update/proposal setting | Use in the present analysis |
|---|---|---|---|
| | | y, beta, the current precision structure, and the previous ystar value as the Gibbs starting point. | parameter in the main diagnostic tables. |
| h.hat / h | Subject-specific kernel effect. | In the present analysis, est.h = FALSE, so h was integrated out in the likelihood used for model fitting and was not sampled in the main MCMC loop. Posterior summaries of h can be computed after fitting when needed. | Not monitored as a stochastic chain parameter in the reported diagnostics. |

## eAppendix 2. Task table and diagnostic settings

eTable 2.1. Exact task-table strata for the 431 *bkmr* model-fitting tasks.

| Bundle | Scenario | Tasks | n | varsel | hfun | beta.true | Iterations | Warmup |
|---|---|---|---|---|---|---|---|---|
| balanced_main | balanced_bkmr_n40_varsel_true | 50 | 40 | TRUE | 2 | 0.5 | 10,000 | 5,000 |
| balanced_main | balanced_bkmr_n80_varsel_false | 50 | 80 | FALSE | 2 | 0.5 | 10,000 | 5,000 |
| balanced_main | balanced_bkmr_n160_varsel_false | 50 | 160 | FALSE | 2 | 0.5 | 10,000 | 5,000 |
| balanced_main | balanced_bkmr_n320_varsel_false | 50 | 320 | FALSE | 2 | 0.5 | 10,000 | 5,000 |
| balanced_ablation | balanced_bkmr_n40_varsel_false | 50 | 40 | FALSE | 2 | 0.5 | 10,000 | 5,000 |
| balanced_longrun_seed3 | balanced_bkmr_n40_varsel_false_longrun | 3 | 40 | FALSE | 2 | 0.5 | 50,000 | 25,000 |
| balanced_longrun_seed3 | balanced_bkmr_n40_varsel_false_longrun | 3 | 40 | FALSE | 2 | 0.5 | 100,000 | 50,000 |
| balanced_longrun_seed3 | balanced_bkmr_n80_varsel_false_longrun | 3 | 80 | FALSE | 2 | 0.5 | 50,000 | 25,000 |
| balanced_longrun_seed3 | balanced_bkmr_n80_varsel_false_longrun | 3 | 80 | FALSE | 2 | 0.5 | 100,000 | 50,000 |
| balanced_n40_longrun_single | balanced_bkmr_n40_single_longrun | 1 | 40 | FALSE | 2 | 0.5 | 50,000 | 25,000 |
| balanced_n40_longrun_single | balanced_bkmr_n40_single_longrun | 1 | 40 | FALSE | 2 | 0.5 | 100,000 | 50,000 |
| default_support | default_bkmr_n40_varsel_true | 20 | 40 | TRUE | 3 | 2.0 | 10,000 | 5,000 |
| default_support | default_bkmr_n80_varsel_false | 20 | 80 | FALSE | 3 | 2.0 | 10,000 | 5,000 |
| public_main | step1_main_exact_varsel_true | 1 | 40 | TRUE | 3 | 2.0 | 10,000 | 5,000 |
| public_main | step3_seed_1_exact_varsel_true | 1 | 40 | TRUE | 3 | 2.0 | 10,000 | 5,000 |
| public_main | step3_seed_2_exact_varsel_true | 1 | 40 | TRUE | 3 | 2.0 | 10,000 | 5,000 |
| public_main | step3_seed_3_exact_varsel_true | 1 | 40 | TRUE | 3 | 2.0 | 10,000 | 5,000 |
| public_main | step4_long_exact_varsel_false | 1 | 40 | FALSE | 3 | 2.0 | 100,000 | 50,000 |
| public_seed10 | public_bkmr_seed10_varsel_true | 10 | 40 | TRUE | 3 | 2.0 | 10,000 | 5,000 |
| public_dataset_sweep | public_bkmr_dataset_sweep_n40_varsel_true | 50 | 40 | TRUE | 3 | 2.0 | 10,000 | 5,000 |
| public_dataset_sweep | public_bkmr_dataset_sweep_n80_varsel_false | 50 | 80 | FALSE | 3 | 2.0 | 10,000 | 5,000 |
| public_n_sweep | public_bkmr_n_sweep_varsel_false | 1 | 40 | FALSE | 3 | 2.0 | 10,000 | 5,000 |
| public_n_sweep | public_bkmr_n_sweep_varsel_false | 1 | 80 | FALSE | 3 | 2.0 | 10,000 | 5,000 |
| public_n_sweep | public_bkmr_n_sweep_varsel_false | 1 | 160 | FALSE | 3 | 2.0 | 10,000 | 5,000 |
| public_n_sweep | public_bkmr_n_sweep_varsel_false | 1 | 320 | FALSE | 3 | 2.0 | 10,000 | 5,000 |
| public_longrun_50k_100k | public_bkmr_longrun_varsel_false | 1 | 40 | FALSE | 3 | 2.0 | 50,000 | 25,000 |
| public_longrun_50k_100k | public_bkmr_longrun_varsel_false | 1 | 40 | FALSE | 3 | 2.0 | 100,000 | 50,000 |
| public_longrun_seed3 | public_bkmr_longrun_seed3_varsel_false | 3 | 40 | FALSE | 3 | 2.0 | 50,000 | 25,000 |
| public_longrun_seed3 | public_bkmr_longrun_seed3_varsel_false | 3 | 40 | FALSE | 3 | 2.0 | 100,000 | 50,000 |

*Note:* Rows were grouped only when bundle, scenario, n, varsel, hfun, beta.true, iterations, and warmup were identical. The 29 strata sum to 431 tasks.

eTable 2.2. Overall task status and diagnostic settings.

| Item | Value |
|---|---|
| Tasks | 431 |
| Completed fits | 430 |
| Numerical non-completions | 1 |
| Diagnostic package | posterior 1.7.0 |
| R-hat definition | maximum of rank-normalized split R-hat and rank-normalized folded-split R-hat |
| R-hat threshold | 1.01 |
| ESS threshold | 400 |

| Initial-value base seed | 20260621 |
|---|---|
| Chain seeds | 142709412; 1107285817; 1907135512; 672667230 |
| Starting values | chain-specific randomized binomial starts: h.hat ~ Uniform(-2, 2), beta ~ Uniform(-1, 1), positive r and \|ystar\| as exp(Uniform(-1, 1)), lambda ~ log-Uniform(1, 20), ystar sign constrained by y, sigsq.eps = 1, delta ~ Bernoulli(0.5) for varsel = TRUE with r = 0 where delta = 0 |

## eAppendix 3. Overall task-level diagnostic results

eTable 3.1. Overall completion and diagnostic pass counts.

| Metric | Count | Comment |
|---|---|---|
| Prespecified tasks | 431 | All tasks in task_table.csv. |
| Fit completed | 430/431 | Returned a *bkmr* fit object. |
| Numerical non-completions | 1/431 | One task returned a numerical fit error in all_task_results.csv. |
| R-hat <= 1.01, considered alone | 55/431 | Primary R-hat threshold. |
| bulk-ESS >= 400, considered alone | 85/431 | Bulk ESS criterion alone. |
| tail-ESS >= 400, considered alone | 44/431 | Tail ESS criterion alone. |
| bulk-ESS and tail-ESS >= 400 | 44/431 | Both ESS criteria. |
| R-hat <= 1.01 plus both ESS criteria | 30/431 | Primary diagnostic pass among prespecified tasks. |
| Primary diagnostic pass among completed fits | 30/430 | Primary pass denominator limited to completed fits. |

## eAppendix 4. Scenario-level diagnostics

eTable 4.1. Balanced generator and n = 40 ablation summaries.

| Scenario | Completed | Median (worst) R-hat | Median (worst) bulk-ESS | Median (worst) tail-ESS | Primary pass |
|---|---|---|---|---|---|
| n=40; varsel=TRUE | 50/50 | 1.043 (1.859) | 116.4 (5.7) | 48.0 (11.4) | 1/50 |
| n=80; varsel=FALSE | 50/50 | 1.035 (2.043) | 126.7 (5.3) | 53.7 (11.1) | 1/50 |
| n=160; varsel=FALSE | 50/50 | 1.013 (1.593) | 417.9 (6.8) | 243.2 (13.4) | 10/50 |
| n=320; varsel=FALSE | 49/50 | 1.012 (1.578) | 461.2 (6.9) | 322.5 (11.5) | 11/50 |
| n=40; varsel=FALSE; ablation | 50/50 | 1.051 (1.586) | 80.9 (6.9) | 44.9 (10.9) | 0/50 |

eTable 4.2. Balanced three-seed long-run diagnostics.

| n | Iterations | Warmup | Completed | Primary pass | Median (worst) R-hat | Median (worst) bulk-ESS | Median (worst) tail-ESS |
|---|---|---|---|---|---|---|---|
| n=40 | 50000 | 25000 | 3/3 | 2/3 | 1.009 (1.011) | 748.6 (365.1) | 797.8 (345.5) |
| n=40 | 100000 | 50000 | 3/3 | 2/3 | 1.006 (1.313) | 785.9 (9.9) | 781.1 (15.3) |
| n=80 | 50000 | 25000 | 3/3 | 1/3 | 1.004 (1.293) | 1002.1 (10.6) | 394.6 (11.5) |
| n=80 | 100000 | 50000 | 3/3 | 2/3 | 1.002 (1.575) | 2552.2 (6.8) | 1010.8 (11.3) |

eTable 4.3. Default-generator and public-verification support summaries by exact task-table stratum.

| Bundle | Exact task-table stratum | Completed | Primary pass | Median (worst) R-hat | Median (worst) bulk-ESS | Median (worst) tail-ESS |
|---|---|---|---|---|---|---|
| default_support | default_bkmr_n40_varsel_true n=40; varsel=TRUE; iter=10,000; warmup=5,000 | 20/20 | 0/20 | 1.670 (2.917) | 6.5 (4.6) | 14.3 (11.4) |
| default_support | default_bkmr_n80_varsel_false n=80; varsel=FALSE; iter=10,000; warmup=5,000 | 20/20 | 0/20 | 1.361 (2.124) | 9.7 (5.3) | 22.6 (11.5) |
| public_main | step1_main_exact_varsel_true n=40; varsel=TRUE; iter=10,000; warmup=5,000 | 1/1 | 0/1 | 1.110 (1.110) | 26.6 (26.6) | 54.4 (54.4) |
| public_main | step3_seed_1_exact_varsel_true n=40; varsel=TRUE; iter=10,000; warmup=5,000 | 1/1 | 0/1 | 1.118 (1.118) | 34.5 (34.5) | 170.9 (170.9) |
| public_main | step3_seed_2_exact_varsel_true n=40; varsel=TRUE; iter=10,000; warmup=5,000 | 1/1 | 0/1 | 1.408 (1.408) | 8.6 (8.6) | 12.5 (12.5) |
| public_main | step3_seed_3_exact_varsel_true n=40; varsel=TRUE; iter=10,000; warmup=5,000 | 1/1 | 0/1 | 1.076 (1.076) | 43.6 (43.6) | 59.8 (59.8) |
| public_main | step4_long_exact_varsel_false n=40; varsel=FALSE; iter=100,000; warmup=50,000 | 1/1 | 0/1 | 1.038 (1.038) | 102.6 (102.6) | 148.9 (148.9) |
| public_seed10 | public_bkmr_seed10_varsel_true n=40; varsel=TRUE; iter=10,000; warmup=5,000 | 10/10 | 0/10 | 1.269 (1.583) | 11.8 (6.9) | 23.9 (11.4) |
| public_dataset_sweep | public_bkmr_dataset_sweep_n40_varsel_true n=40; varsel=TRUE; iter=10,000; warmup=5,000 | 50/50 | 0/50 | 1.633 (3.229) | 6.6 (4.4) | 17.6 (11.3) |

| Bundle | Exact task-table stratum | Completed | Primary pass | Median (worst) R-hat | Median (worst) bulk-ESS | Median (worst) tail-ESS |
|---|---|---|---|---|---|---|
| public_dataset_sweep | public_bkmr_dataset_sweep_n80_varsel_false n=80; varsel=FALSE; iter=10,000; warmup=5,000 | 50/50 | 0/50 | 1.458 (2.541) | 8.1 (4.8) | 27.2 (11.2) |
| public_n_sweep | public_bkmr_n_sweep_varsel_false n=40; varsel=FALSE; iter=10,000; warmup=5,000 | 1/1 | 0/1 | 1.313 (1.313) | 10.6 (10.6) | 105.3 (105.3) |
| public_n_sweep | public_bkmr_n_sweep_varsel_false n=80; varsel=FALSE; iter=10,000; warmup=5,000 | 1/1 | 0/1 | 1.295 (1.295) | 11.0 (11.0) | 26.5 (26.5) |
| public_n_sweep | public_bkmr_n_sweep_varsel_false n=160; varsel=FALSE; iter=10,000; warmup=5,000 | 1/1 | 0/1 | 1.755 (1.755) | 6.0 (6.0) | 17.9 (17.9) |
| public_n_sweep | public_bkmr_n_sweep_varsel_false n=320; varsel=FALSE; iter=10,000; warmup=5,000 | 1/1 | 0/1 | 1.426 (1.426) | 8.1 (8.1) | 24.4 (24.4) |
| public_longrun_50k_100k | public_bkmr_longrun_varsel_false n=40; varsel=FALSE; iter=50,000; warmup=25,000 | 1/1 | 0/1 | 1.071 (1.071) | 49.9 (49.9) | 56.3 (56.3) |
| public_longrun_50k_100k | public_bkmr_longrun_varsel_false n=40; varsel=FALSE; iter=100,000; warmup=50,000 | 1/1 | 0/1 | 1.043 (1.043) | 95.3 (95.3) | 92.5 (92.5) |
| public_longrun_seed3 | public_bkmr_longrun_seed3_varsel_false n=40; varsel=FALSE; iter=50,000; warmup=25,000 | 3/3 | 0/3 | 1.100 (1.115) | 34.3 (28.3) | 63.3 (49.5) |

| Bundle | Exact task-table stratum | Completed | Primary pass | Median (worst) R-hat | Median (worst) bulk-ESS | Median (worst) tail-ESS |
|---|---|---|---|---|---|---|
| public_longrun_seed3 | public_bkmr_longrun_seed3_varsel_false<br>n=40; varsel=FALSE;<br>iter=100,000; warmup=50,000 | 3/3 | 0/3 | 1.040 (1.057) | 126.9 (91.2) | 139.7 (49.0) |

Note: This table includes all 40 default_support tasks and all 127 tasks in the public-verification bundles (167 tasks total). Rows were defined by bundle, scenario, n, varsel, iterations, and warmup; settings that differed in sample size or chain length were not pooled. Primary pass required maximum rank-normalized R-hat <= 1.01 and both bulk-ESS and tail-ESS >= 400.

## eAppendix 5. Parameter-group convergence diagnostics

The reproduction outputs include one parameter-level diagnostic CSV for each of the 430 completed fits. To identify which parameter groups contributed to diagnostic concerns, we summarized rank-normalized R-hat, bulk-ESS, and tail-ESS by parameter group. A diagnostic flag was defined for a posterior-parameter row as R-hat > 1.01, bulk-ESS < 400, or tail-ESS < 400. beta and lambda each contributed one posterior-parameter row per completed fit, and both frequently showed elevated R-hat or low bulk-ESS or tail-ESS. r contributed four rows per completed fit because M = 4 exposure components were monitored. Although the median r diagnostics were acceptable, many tasks had at least one flagged r parameter, indicating that diagnostic concerns were concentrated in subsets of r parameters or tasks rather than being apparent from the median alone. Variable-selection behavior was summarized separately: rdelta_move is an acceptance-only variable-selection move summary and is not directly comparable with posterior-parameter R-hat or ESS. In the binomial/probit model, sigsq.eps was fixed at 1, so R-hat and ESS were undefined by construction.

eTable 5.1. Parameter-group diagnostic summary across the 430 completed fits.

| Parameter group | Rows / tasks | Rows with R-hat / ESS | R-hat median (max) | bulk-ESS median (min) | tail-ESS median (min) | Parameter rows with diagnostic flag | Tasks with any diagnostic flag |
|---|---|---|---|---|---|---|---|
| beta | 430 rows / 430 tasks | 430 | 1.048 (3.229) | 81.4 (4.4) | 92.1 (11.2) | 280/430 | 280/430 |
| lambda | 430 rows / 430 tasks | 430 | 1.026 (1.550) | 168.4 (7.1) | 286.4 (19.2) | 290/430 | 290/430 |
| r | 1,720 rows / 430 tasks | 1,720 | 1.009 (2.043) | 504.4 (5.3) | 443.8 (10.9) | 987/1,720 | 386/430 |
| rdelta_move | 134 move summaries / 134 variable-selection tasks | NA | NA | NA | NA | NA | NA |
| sigsq.eps | 430 rows / 430 tasks | NA | NA | NA | NA | NA | NA |

Note: Rows are parameter-level rows, not tasks. beta and lambda each contributed one row per completed fit. r contributed four rows per completed fit because M = 4 exposure components were monitored. A diagnostic flag was defined as rank-normalized R-hat > 1.01, bulk-ESS < 400, or tail-ESS < 400. The tasks-with-any-diagnostic-flag column counts completed tasks with at least one flagged row in the corresponding parameter group. NA indicates that R-hat and ESS were not applicable: rdelta_move is an acceptance-only variable-selection move summary, not a posterior parameter, and sigsq.eps was fixed at 1 in the binomial/probit model.